\documentclass[twocolumn,aps,showpacs,floatfix,prc]{revtex4}
\usepackage[dvips]{epsfig}
\begin{document}
%\draft
\title{New thermonuclear reaction rate equations for radiative neutron capture}

\author{ Vinay Singh$^{1\S\dagger}$, Joydev Lahiri$^{2\S}$ and D. N. Basu$^{3\S\dagger}$}

\affiliation{$^{\S}$Variable Energy Cyclotron Centre, 1/AF Bidhan Nagar, Kolkata 700064, INDIA}
\affiliation{$^\dagger$Homi Bhabha National Institute, Training School Complex, Anushakti Nagar, Mumbai 400085}

\email[E-mail 1: ]{vsingh@vecc.gov.in}
\email[E-mail 2: ]{joy@vecc.gov.in}
\email[E-mail 3: ]{dnb@vecc.gov.in} 

\date{\today }

\begin{abstract}

    The radiative neutron capture reaction rates have been studied at very low energies which are of interest for nuclear astrophysics. The rates for many of these reactions have remained independent of temperature so far. The temperature dependence of the thermonuclear reaction rates have been explored within the statistical model. Apart from the compound nuclear contribution, the pre-equilibrium as well as the direct effects have been taken into account. The corresponding Maxwellian-averaged thermonuclear reaction rates of relevance in astrophysical plasmas at temperatures in the range from 10$^6$ K to 10$^{10}$ K have been calculated. Analytical expression as a function of $T_9$ has been provided for $^6$Li(n,$\gamma$)$^7$Li by fitting the calculated reaction rate.
\vskip 0.2cm

\end{abstract}

\maketitle

\noindent
\section{Introduction}
\label{section1}

    The astrophysical thermonuclear reactions generate energy that makes stars shine. These are also responsible for the synthesis of the elements in stars. The interstellar medium is enriched with the nuclear ashes when stars eject part of their matter through various means. These processes provide the building blocks for the birth of new stars, of planets and of life itself. The theory of production of elements is called nucleosynthesis and in describing the nuclear processes in stars that are located so far away from us in space and time, it is remarkably successful. The process of synthesis of the elements can be broadly classified into two categories: the primordial or big-bang nucleosynthesis and the stellar nucleosynthesis. As the name suggests the primordial nucleosynthesis refers to what happened at the beginning of the universe when light elements such as D, T, $^{3,4}$He, $^{6,7}$Li and $^7$Be were synthesized while stellar nucleosynthesis occurs in stars and causes synthesis of heavier elements. It is also worthy of attention how the theory predicts these processes based on the quantum mechanical properties of atomic nuclei. The nuclear energy generation in stars, nucleosynthesis and other issues at the intersection of astrophysics and nuclear physics make up the science of nuclear astrophysics. Similar to most areas of physics, it involves both experimental and theoretical activities. 

    The nuclear reaction cross sections and its convolution with Maxwell-Boltzmann distribution of energies are important for modeling many physical phenomena occurring under extreme conditions \cite{Bu57,Fo64,Cl83}. Such environments of very high temperature or density exist in main-sequence stars and compact stars which are in final stages of their evolutionary development. The exothermic nuclear fusion drives nuclear explosions in the surface layers of the accreting white dwarfs (nova events), in the cores of massive accreting white dwarfs (type Ia supernovae) \cite{Ni97,Ho06} and in the surface layers of accreting neutron stars (type I X-ray bursts and superbursts \cite{St06,Sc03,Cu06,Gu07}). All these astrophysical processes require precise knowledge of nuclear reaction rates obtained by convoluting cross sections with Maxwell-Boltzmann distribution of energies. The Maxwellian-averaged thermonuclear reaction rate per particle pair $<\sigma v>$ at temperature $T$, is given by the following integral \cite{Ad11,Fo67,Bo08}:

\begin{equation}
 <\sigma v> = \Big[\frac{8}{\pi m (k T)^3 } \Big]^{1/2} \int \sigma(E) E \exp(-E/k T) dE,
\label{seqn1}
\end{equation}
\noindent
where $v$ is the relative velocity, $k$ and $m$ are the Boltzmann constant and the reduced mass of the reacting nuclei, respectively. Therefore, the reaction rate between two nuclei can be written as $r_{12}=\frac{n_1n_2}{1+\delta_{12}}<\sigma v>$ where $n_1$ and $n_2$ are the number densities of nuclei of types 1 and 2. The Kronecker delta $\delta_{12}$ prevents double counting in the case of identical particles.
    
    Several works \cite{Fo88,Ma89,Sm93,An99,An04} regarding reaction rates have been done in past. Except a few neutron induced reactions, all other reaction rates have temperature dependences. At thermal energies neutron absorption cross section shows an approximate $1/v$ behavior. Hence, using $\sigma(E) \propto E^{-1/2}$ in Eq.(1) immediately shows that the reaction rates are approximately constant with respect to temperature at low energies. However, this fact is true for thermal neutrons ($\sim$ 0.025 eV) only with energies of the order of eV and below. But at energies of astrophysical interest, the neutron induced reaction cross sections can be given by $\sigma(E)=\frac{R(E)}{v}$ \cite{Bl55}, where $R(E)$ is a slowly varying function of energy \cite{Mu10} and is similar to the astrophysical S-factor and one expects $\langle\sigma v\rangle$ to be dependent on temperature. Since $^6$Li(n,$\gamma$)$^7$Li, $^{10}$B(n,$\gamma$)$^{11}$B, $^{12}$C(n,$\gamma$)$^{13}$C and $^{14}$N(n,$\gamma$)$^{15}$N are the reactions with rates independent of temperature, it attracted special attention for further investigation of temperature dependence. 

\vspace{0.0cm}
\noindent
\section{Theoretical formalism}
\label{section2}
\vspace{0.0cm}

    The thermonuclear reaction rates can be obtained by convoluting fusion cross sections with Maxwell-Boltzmann distribution of energies. These cross sections can vary by several orders of magnitude across the required energy range. The low energy fusion cross sections $\sigma$, some of which are not sufficiently well known, can be obtained from laboratory experiments. However, there are cases, in particular involving the weak interaction such as the basic p$+$ p fusion to deuterium in the solar p-p chain, where no experimental data are available and one completely relies on theoretical calculations \cite{Ad11}. The theoretical estimates of the thermonuclear reaction rates depend on the various approximations used. Several factors influence the measured values of the cross sections. We need to account for the Maxwellian-averaged thermonuclear reaction rates in the network calculations used in primordial and stellar nucleosynthesis.

    The reaction rates used in the Big Bang Nucleosynthesis (BBN) reaction network have temperature dependences except $^6$Li(n,$\gamma$)$^7$Li, $^{10}$B(n,$\gamma$)$^{11}$B, $^{12}$C(n,$\gamma$)$^{13}$C and $^{14}$N(n,$\gamma$)$^{15}$N which are constant with respect to temperature. The computer code TALYS \cite{Talys} allows a comprehensive astrophysical reaction rate calculations apart from other nuclear physics calculations. To a good approximation, in the interior of stars the assumption of a thermodynamic equilibrium holds and nuclei exist both in the ground and excited states. This assumption along with cross sections calculated from compound nucleus model for various excited states facilitates Maxwellian-averaged reaction rates. For stellar evolution models this is quite an important input. The nuclear reaction rates are generally evaluated using the statistical model \cite{Rauscher97,Rauscher10} and astrophysical calculations mostly use these reaction rates. Stellar reaction rate calculations have been routinely done in past \cite{Rauscher00,Rauscher01}. However, TALYS has extended these Hauser-Feshbach statistical model \cite{Ha52} calculations by adding some new and important features. Apart from coherent inclusion of fission channel it also includes reaction mechanism that occurs before equilibrium is reached, multi-particle emission, competition among all open channels, width fluctuation corrections in detail, coupled channel description in case of deformed nuclei and level densities that are parity-dependent. The nuclear models are also normalized for available experimental data using separate approaches such as on photo-absorption data, the E1 resonance strength or on s-wave spacings, the level densities.

    In the low energy domain, compound nucleus is formed by the fusion of the projectile and the target nuclei. While the total energy $E^{tot}$ is fixed from energy conservation, the total spin $J$ and parity $\Pi$ can have a range of values. The reaction obeys the following conservation laws,
$$
E_{a}+S_{a}\ =\ E_{a'}+E_{x}+S_{a'}=E^{tot},~~~~{\rm energy~conservation},  
$$
$$
s+I+l\ =\ s'+I'+l'=J,~~~~{\rm angular~momentum~conservation},
$$
\begin{center}
  $\pi_{0}\Pi_{0}(-1)^{l}\ = \pi_f\Pi_f(-1)^{l'}=\Pi,~~~~{\rm parity~conservation}.$
\end{center}
The formula for binary cross section, assuming the compound nucleus model, is given by

\begin{eqnarray}
\sigma_{\alpha\alpha'}^{comp}\ &&=\ D^{comp}\frac{\pi}{k^{2}}\sum_{J=mod(I+s,1)}^{l_{\max}+I+s}
\sum_{\Pi=-1}^{1}\frac{2J+1}{(2I+1)(2s+1)}  \nonumber\\
&&\sum_{j=|J-I|}^{J+I}\sum_{l=|j-s|}^{j+s}\sum_{j'=|J-I'|}^{J+I'}\sum_{l'=|j'-s'|}^{j'+s'} \delta_{\pi}(\alpha)\delta_{\pi}(\alpha')
\end{eqnarray}

\begin{center}
   $\displaystyle \times\ \frac{T_{\alpha lj}^{J}(E_{a})\langle T_{\alpha' l'j'}^{J}(E_{a'})\rangle}{\sum_{\alpha'',l'',j''}\delta_{\pi}(\alpha'')\langle T_{\alpha''l''j''}^{J}(E_{a''})\rangle}W_{\alpha lj\alpha'l'j'}^{J}$
\end{center}
\noindent
where

$E_{a}=$ the energy of the projectile 

$l=$ the orbital angular momentum of the projectile

$s=$ the spin of the projectile

$j=$ the total angular momentum of the projectile

$\pi_{0}=$ the parity of the projectile

$\delta_{\pi}(\alpha)=\left\{ \begin{array}{ll}
1 ~~&{\rm if}~~(-1)^{l}\pi_{0}\Pi_{0}=\Pi \\ 
 0~~& {\rm otherwise}
\end{array}\right.$ 

$\alpha=$ the designation of the channel for the initial projectile-target system:

$\alpha=\{a,\ s,\ E_{a},\ E_{x}^{0},\ I,\ \Pi_{0}\}$, where $a$ and $E_{x}^{0}$ are the type of the projectile and  the excitation energy (which is zero usually) of the target nucleus, respectively 

$l_{\max}=$ the maximum l-value of the projectile

$S_{a}=$ the separation energy

$E_{a'} =$  the energy of the ejectile

$l'=$ the orbital angular momentum of the ejectile

$s'=$ the spin of the ejectile

$j'=$ the total angular momentum of the ejectile 

$\pi_{f}=$ the parity of the ejectile

$\delta_{\pi}(\alpha')=\left\{ \begin{array}{ll}
1 ~~&{\rm if}~~(-1)^{l'}\pi_{f}\Pi_{f}=\Pi \\ 
 0~~& {\rm otherwise}
\end{array}\right.$ 

$\alpha'=$ the designation of channel for the ejectile-residual nucleus final system:

$\alpha'=\{a',\ s',\ E_{a'},\ E_{x},\ I',\ \Pi_f\}$, where $a'$ and  $E_{x}$ are the type of the ejectile and the residual nucleus excitation energy, respectively

$I=$ the spin of target nucleus

$\Pi_{0}=$ the parity of target nucleus

$I'=$ the spin of residual nucleus

$\Pi_f=$ the parity of residual nucleus

$J=$ the total angular momentum of the compound system

$\Pi=$ the parity of the compound system

$D^{comp}=$ the depletion factor so as to take into account for pre-equilibrium and direct effects

$k=$ the wave number of the relative motion

$T=$ the transmission coefficient

$W=$ the correction factor for width fluctuation (WFC).

    The velocities of both the targets and projectiles obey Maxwell- Boltzmann distributions corresponding to ionic plasma temperature $T$ at the site. The astrophysical nuclear reaction rate can be calculated by folding the Maxwell-Boltzmann energy distribution for energies $E$ at the given temperature $T$ with the cross section given by Eq.(2). Additionally, target nuclei exist both in ground and excited states. The relative populations of various energy states of nuclei with excitation energies $E_{x}^{\mu}$ and spins $I^{\mu}$ in thermodynamic equilibrium follows the Maxwell-Boltzmann distribution. In order to distinguish between different excited states the superscript $\mu$ is used along with the incident $\alpha$ channel in the formulas that follow. Taking due account of various target nuclei excited state contributions, the effective nuclear reaction rate in the entrance channel $\alpha\rightarrow\alpha'$ can be finally expressed as 
    
\begin{equation}
 N_{A}\langle\sigma v\rangle_{\alpha\alpha'}^{*}(T)=\left(\frac{8}{\pi m}\right)^{1/2}\frac{N_{A}}{(kT)^{3/2}G(T)}\times\ 
\end{equation}

\begin{center}
   $\displaystyle  \int_{0}^{\infty}\sum_{\mu}\frac{(2I^{\mu}+1)}{(2I^{0}+1)}\sigma_{\alpha\alpha'}^{\mu} (E)E\exp\left(-\frac{E+E_{x}^{\mu}}{kT}\right)dE,$
\end{center}
where $N_{A}$ is the Avogadro number which is equal to 6.023$\times 10^{23}$, $k$ and $m$ are the Boltzmann constant and the reduced mass in the $\alpha$ channel, respectively, and
 
\begin{center}
  $G(T)=\displaystyle \sum_{\mu}(2I^{\mu}+1)/(2I^{0}+1)\exp(-E_{x}^{\mu}/kT)$
\end{center}
is the temperature dependent normalized partition function. By making use of the reciprocity theorem \cite{Ho76}, the reverse reaction cross sections or rates can also be estimated. 
      
\noindent
\section{Calculations and results}
\label{section3}

    All the reaction rates used in the BBN reaction network have temperature dependences except a few such as $^6$Li(n,$\gamma$)$^7$Li, $^{10}$B(n,$\gamma$)$^{11}$B, $^{12}$C(n,$\gamma$)$^{13}$C and $^{14}$N(n,$\gamma$)$^{15}$N. The experimental status is that only one direct measurement of the $^6$Li(n,$\gamma$)$^7$Li cross sections has been performed \cite{To00} at stellar energies. The Malaney-Fowler reaction rate \cite{Ma89} for $^6$Li(n,$\gamma$)$^7$Li in the BBN reaction network calculations has been taken as 5.10$\times$10$^3$ cm$^3$s$^{-1}$mol$^{-1}$ which is constant with respect to temperature. In a recent calculation \cite{Su10} it was found to be (8.5 $\pm$ 1.7)$\times$10$^3$ cm$^3$s$^{-1}$mol$^{-1}$ where the error resulted from the uncertainties of spectroscopic factors and scattering potential depth. The astrophysical reaction rate was found to be higher by a factor of 1.7 than the value adopted in previous reaction network calculations \cite{No97,No00,Si19}. For the reactions $^{10}$B(n,$\gamma$)$^{11}$B, $^{12}$C(n,$\gamma$)$^{13}$C and $^{14}$N(n,$\gamma$)$^{15}$N, rates constant with respect to temperature given by $6.62 \times 10^4$, $4.50 \times 10^2$ and $9.94 \times 10^3$ cm$^3$s$^{-1}$mol$^{-1}$, respectively \cite{Wa69}, have been used.
    
    The radiative neutron capture cross section varies inversely as velocity in the range of thermal energies. At these energies, the feature of $\sigma(E) \propto E^{-1/2}$ leads to approximate constancy of thermonuclear reaction rates with respect to plasma temperature. However, above thermal energies, especially in the domain of astrophysics, the neutron induced reaction cross sections deviates from the $1/v$ law and rather follows the dependence given by $\sigma(E)=\frac{R(E)}{v}$ \cite{Bl55}. Thus it is expected that $\langle\sigma v\rangle$ has to have a temperature dependence. As discussed earlier all the thermonuclear reaction rates do have temperature dependences except a few neutron induced reactions. These rates have remained independent of temperature so far and, therefore, attracted special attention for further investigation of their temperature dependences. 

    Due to the reason that the cross sections involved in nuclear astrophysics are very low, one of the main problems is to extrapolate the available data down to very low energies. For this purpose various models, such as the potential model or microscopic approaches, are widely used. These are in general not very flexible to account for the data with reasonably good accuracy. Obviously, polynomial approximation is the simplest way to extrapolate these data \cite{Sm93}. Usually this is used to investigate electron screening effects, where the cross section between bare nuclei is derived from extrapolating the high energy data by a polynomial approximation. Although very simple, the polynomial approximation is not based on a rigorous treatment of the energy dependence of the cross section, and may introduce significant inaccuracies. As mentioned in the previous section, we use here a more rigorous approach.
      
\begin{table}[h!]
\vspace{0.0cm}
\centering
\caption{\label{tab:table1} Reaction Rate in units of cm$^3$s$^{-1}$mol$^{-1}$ for the reaction $^6$Li(n,$\gamma$)$^7$Li as a function of temperature $T_9$ (expressed in units of $10^9$ K) generated from TALYS.}
\vspace{0.2cm}
\begin{tabular}{|l|c|c|c|c|c|}
\hline
\hline
$T_9$&Reaction Rate&$T_9$&Reaction Rate&$T_9$&Reaction Rate \\ \hline

0.0001  	&408.43 &0.3 	&436.44 &2.5 	&824.89  \\
0.0005  	&440.24 &0.4 	&466.02 &3.0 	&883.00  \\
0.001 	&423.09 &0.5 	&492.41 &3.5 	&936.22  \\
0.005 	&414.78 &0.6 	&516.58 &4.0 	&985.55  \\
0.01 	&392.78 &0.7 	&539.18 &5.0 	&1074.75 \\
0.05  	&369.81 &0.8 	&560.55 &6.0 	&1153.17  \\
0.1 	&375.42 &0.9 	&580.88 &7.0 	&1222.50  \\
0.15 	&388.01 &1.0 	&600.31 &8.0 	&1284.74  \\
0.2 &403.71 &1.5	  &686.76 &9.0 	&1342.36  \\
0.25 &420.29 &2.0	  &760.28 &10.0 	&1397.99 \\

\hline
\hline
\end{tabular} 
\vspace{0.0cm}
\end{table}
\noindent 

\begin{figure}[ht]
\vspace{0.0cm}
\eject\centerline{\epsfig{file=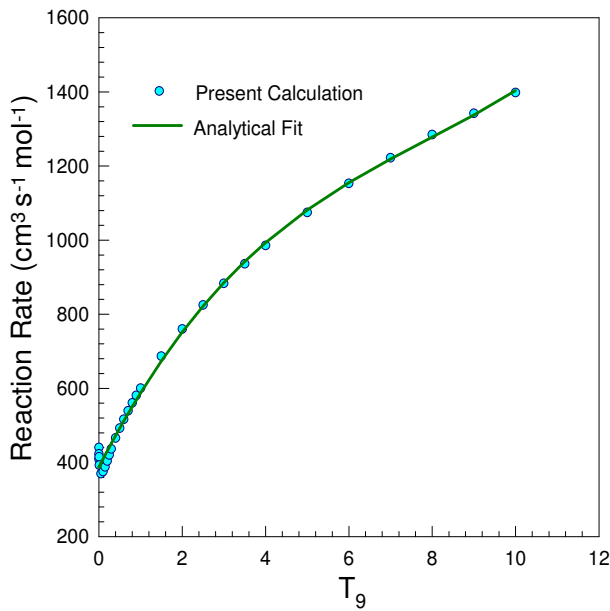,height=8cm,width=8cm}}
\caption{(Color online) Plot of reaction rate as function of temperature $T_9$. The dots represent results of the present calculations while the continuous line represents the fit to it.}
\label{fig1}
\vspace{0.0cm}
\end{figure}

    The reaction rates for the reactions $^6$Li(n,$\gamma$)$^7$Li, $^{10}$B(n,$\gamma$)$^{11}$B, $^{12}$C(n,$\gamma$)$^{13}$C and $^{14}$N(n,$\gamma$)$^{15}$N have been calculated theoretically using the TALYS \cite{Talys} code. Although, in these reactions the nuclei involved are light, the results of the calculations are expected to be reasonable, particularly as the reaction being induced by low energy neutrons and not by charged particles \cite{An99} implying dominant contributions from compound nuclear reaction. Moreover, apart from the compound nuclear contribution, TALYS accounts for the pre-equilibrium and the direct effects as well. However, the pre-equilibrium effects do not play any role for $^6$Li(n,$\gamma$)$^7$Li at energies below the $^7$Li neutron separation energy of 7.25 MeV. The results of the calculations for $^6$Li(n,$\gamma$)$^7$Li reaction rate in units of cm$^3$s$^{-1}$mol$^{-1}$ as a function of temperature $T_9$ generated from TALYS code have been presented in Table-1 whereas for the rest of the reactions, $^{10}$B(n,$\gamma$)$^{11}$B, $^{12}$C(n,$\gamma$)$^{13}$C and $^{14}$N(n,$\gamma$)$^{15}$N, rates have been found to be constant given by $5.3318 \times 10^3 \pm 8.67$, $2.8749 \times 10^3 \pm 5.81$ and $4.2081 \times 10^4 \pm 3.97 \times 10^1$ cm$^3$s$^{-1}$mol$^{-1}$, respectively.
    
\noindent
\section{Analytical Parametrization of Reaction Rates}
\label{section4}

    In order to provide analytical parametrization of reaction rate, the results presented in Table-1 has been fitted quite accurately as a function of $T_9$. The parameterization adopted for this neutron induced reaction is  
    
\begin{equation}
  N_{A}\langle\sigma v\rangle(T) = a_0\left( 1 + \sum_{i=1}^N a_i T_9^i \right)
\label{seqn4}
\end{equation}
\noindent
which provides excellent fit to the calculated values as evident from Fig.-1. The values of the parameters obtained are $a_0=387.75 \pm 0.95$, $a_1=0.5659 \pm 0.0059$, $a_2=(-0.52746 \pm 0.01586)\times 10^{-1}$ and $a_3=(0.22354 \pm 0.01168)\times 10^{-2}$. The errors in the fitted parameters are calculated from the correlation matrix in the final stage of the fitting procedure when changes in the fitted parameters by amounts equal to the corresponding uncertainties in the fitted parameters cause changes in the corresponding quantity by less than the stipulated value. Thus large uncertainty in a fitted parameter implies that the hyper-surface is rather flat with respect to that parameter. We also find that the contributions from terms containing higher orders of $T_9$ are insignificant. Hence, terms only up to third order in $T_9$ has been retained in Eq.(4). Also any other form such as that provided in \cite{An04} for charged particle induced reactions could not be fitted with even a reasonable value of chi-square per degrees of freedom. 

    The plot of reaction rate as a function of temperature $T_9$ is shown in Fig.-1. The dots represent results of the present calculations provided in Table-I while the continuous line represents its fitting by the function of $T_9$: $387.75\times \left[ 1+0.5659 T_9-0.052746T_9^2+0.0022354T_9^3\right]$. This yields a new reaction rate equation given by 

\begin{eqnarray}
N_A<\sigma v> =(387.75 \pm 0.95)+(219.45\pm 1.91)T_9 \nonumber\\
-(20.45\pm 0.59)T_9^2+(0.867\pm 0.045)T_9^3  
\label{seqn5}
\end{eqnarray}
\noindent
expressed in units of cm$^3$s$^{-1}$mol$^{-1}$. This reaction rate is meant to supersede the earlier reaction rate used in the BBN calculations.

\noindent
\section{Summary and conclusion}
\label{section5}
    
    A new analytical expression for the thermonuclear reaction rate of $^6$Li(n,$\gamma$)$^7$Li neutron capture reaction has been developed as a function of $T_9$. This has been achieved by fitting the results of the reaction rate generated from nuclear reaction theory calculations. Apart from the compound nuclear contribution, the pre-equilibrium and the direct effects have been accounted for. The thermonuclear reaction rate for $^6$Li(n,$\gamma$)$^7$Li has been found to be temperature dependent which increases monotonically beyond thermal neutron energies up to 10$T_9$ that corresponds to $\sim$ 0.86 MeV while those for $^{10}$B(n,$\gamma$)$^{11}$B, $^{12}$C(n,$\gamma$)$^{13}$C and $^{14}$N(n,$\gamma$)$^{15}$N remained constant within the same range of temperatures. Such a result implies that compound nucleus picture is more or less true representations of slightly heavier nuclei such as $^{11}$B, $^{13}$C and $^{15}$N. These new reaction rates have been incorporated in the Kawano/Wagoner BBN code \cite{Wa69,Wa67,Ka92} modified by Singh et al. \cite{Si19}. The primordial elemental abundances, however, remained unchanged when these new reaction rates have been used for it. This is expected as these reaction do not play any significant role in standard Big Bang nucleosynthesis as there are little free
neutrons available. Also in neutron-rich nucleosynthesis the rate is too slow to compete with other reactions. Nevertheless, these new reaction rates may find usefulness in other domains of nuclear astrophysics such as stellar burning and stellar nucleosynthesis.

\end{document}